\documentclass{aa}
\usepackage[T1]{fontenc}
\usepackage{ae,aecompl}
\usepackage{natbib}
\usepackage{graphicx}	
\usepackage{amsmath}	
\usepackage{amssymb}	

\newcommand{\bfy}{{\bf y}}
\newcommand{\bfn}{{\bf n}}
\newcommand{\bfs}{{\bf s}}
\newcommand{\bfq}{{\bf q}}
\newcommand{\ro}{\rho_{\rm out}}
\newcommand{\ri}{\rho_{\rm inp}}

\begin{document} 

   \title{Limitations on the recovery of the true AGN variability parameters using Damped Random Walk modeling}

   \titlerunning{Limitations on the recovery of the true AGN variability parameters using DRW modeling}
   \authorrunning{S. Koz{\l}owski}

   \author{Szymon Koz{\l}owski
          \inst{1}
          }

   \institute{Warsaw University Observatory,\\
              Al. Ujazdowskie 4, 00-478 Warszawa, Poland\\
              \email{simkoz@astrouw.edu.pl}
             }

   \date{Received October 13, 2016; accepted November 23, 2016}

\abstract
{The damped random walk (DRW) stochastic process is nowadays frequently used to model aperiodic light curves of active galactic nuclei (AGNs).
A number of correlations between the DRW model parameters, the signal decorrelation timescale and amplitude, and the physical AGN parameters such as the black hole mass or luminosity
have been reported.}
{We are interested in whether it is plausible to correctly measure the DRW parameters from a typical ground-based survey, in particular 
how accurate the recovered DRW parameters are compared to the input ones.}
{By means of Monte Carlo simulations of AGN light curves, we study the impact of the light curve length, the source magnitude (so the photometric properties of a survey),
cadence, and additional light (e.g., from a host galaxy) on the DRW model parameters.}
{The most significant finding is that currently existing surveys are going to return unconstrained DRW
decorrelation timescales, because typical rest-frame data do not probe long enough timescales or 
the white noise part of the power spectral density for DRW.
The experiment length must be at least ten times longer than the true DRW decorrelation timescale, 
being presumably in the vicinity of one year, so meaning the necessity of a minimum 10-years-long AGN light curves (rest-frame).
The DRW timescales for sufficiently long light curves are typically weakly biased, and the exact bias depends on the fitting method and used priors.
The DRW amplitude is mostly affected by the photometric noise (so the source magnitude or the signal-to-noise ratio), cadence, and the AGN host light.}
{Because the DRW parameters appear to be incorrectly determined from typically existing data, 
the reported correlations of the DRW variability and physical AGN parameters from other works seem unlikely to be correct.
In particular, the anti-correlation of the DRW decorrelation timescale with redshift is a manifestation of the survey length being too short.
Application of DRW to modeling typical AGN optical light curves is questioned.}

\keywords{accretion, accretion disks -- galaxies: active -- methods: data analysis -- quasars: general}

\maketitle



\section{Introduction}

Active galactic nuclei (AGNs) variability studies have already entered into a new era. 
This is primarily due to a gigantic increase in data volume from the already existing and forthcoming deep, large sky surveys, but also
due to introduction of new methods that enabled a direct modeling of aperiodic AGN light curves.
They are nowadays routinely modeled using the damped random walk (DRW) model 
(\citealt{2009ApJ...698..895K,2010ApJ...708..927K,2010ApJ...721.1014M,2011ApJ...728...26M,2012ApJ...753..106M},
\citealt{2011AJ....141...93B,2012ApJ...760...51R,2011ApJ...735...80Z,2013ApJ...765..106Z,2016ApJ...819..122Z}),
with potentially some departures as seen both in steeper than DRW power spectral distributions (PSDs) of AGNs from the {\it Kepler} mission (\citealt{2011ApJ...743L..12M,2015MNRAS.451.4328K}), 
steeper than DRW PSDs for more massive Pan-STARRS (\citealt{2016A&A...585A.129S}) and PTF/iPTF AGNs (\citealt{2016arXiv161103082C}),
and from steeper than DRW structure functions (SFs) for luminous SDSS AGNs (\citealt{2016ApJ...826..118K}). 
Modeling AGN light curves using DRW that seem to have other than DRW covariance matrix of the signal is perfectly doable, however at a price of obtaining biased DRW parameters 
(\citealt{2016MNRAS.459.2787K}). \citealt{2014ApJ...788...33K} considered recently a broader class of continuous-time autoregressive moving average (CARMA) models of which DRW is the simplest variant.

An AGN light curve can be described (and fitted) with DRW having just two model parameters: the damping timescale $\tau$ (the signal decorrelation timescale) and the modified variability 
amplitude $\hat{\sigma}$ (or equivalently SF$_\infty=\hat{\sigma}\sqrt{\tau}$; \citealt{2010ApJ...721.1014M}). 
\cite{2009ApJ...698..895K} reported on correlations between the DRW model parameters and physical properties of AGNs.
Using $\sim$9000 SDSS quasars, \citealt{2010ApJ...721.1014M} analyzed these correlations in detail and showed that these two model 
parameters are correlated with the wavelength, luminosity and/or black hole mass.  Or are they?

\begin{figure*}
\centering
\includegraphics[width=14.0cm]{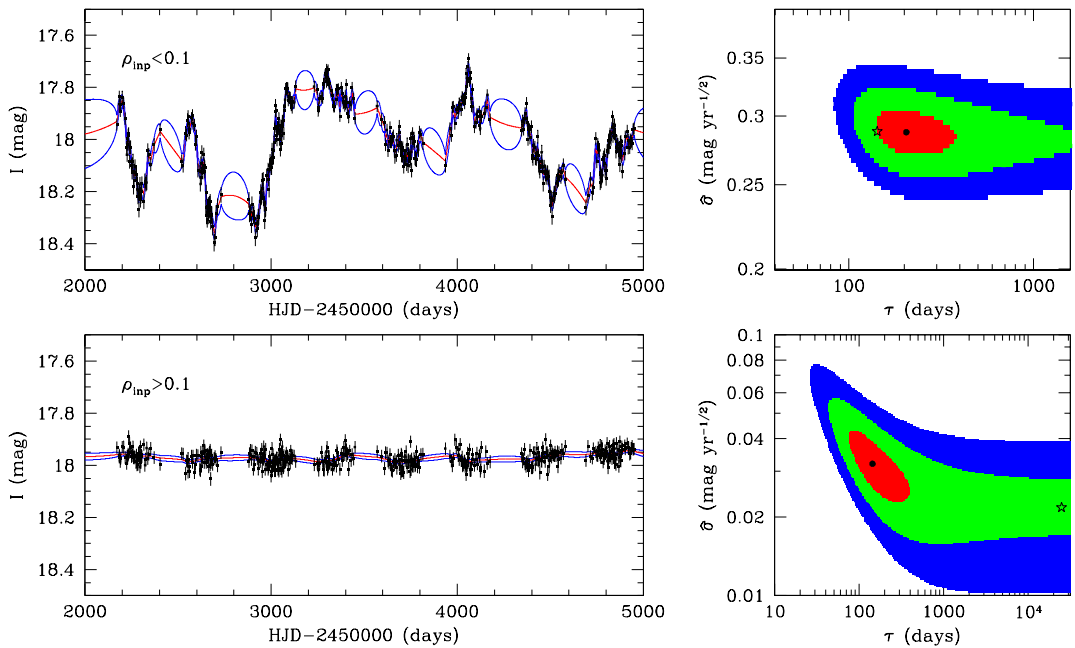}
\caption{Left column presents two examples of simulated AGN light curves for OGLE-III. Both light curves have the mean magnitude $I=18$~mag, $SF_\infty=0.20$ mag, 
but the input timescale is $\tau_{\rm inp}=150$ days (top) and 25000 days (bottom). The experiment lasted approximately $t_{\rm exp}=2800$~days, hence the ratio of the 
decorrelation timescale to the experiment length $\ri=\tau_{\rm inp}t_{\rm exp}^{-1}$ is 0.05 and 9.0, respectively. 
The red line is the best-fit DRW model and the blue lines present $1\sigma$ ``error snakes''. The right column presents the model likelihood surface
for the two light curves. The black point is the maximum likelihood point, while the red, green, and blue areas represent $\Delta\ln{\mathcal{L}}=-0.5$, $-2.0$, and $-4.5$,
corresponding to $\Delta\chi^2=1$, 4, and 9, (or 1, 2, and 3$\sigma$), respectively. The open star marks the input parameters.
The bias in the measured parameters is obvious in both cases (dots and stars do not match).}
\label{fig:lcs}
\end{figure*}

This paper originates from a curiosity of to why cutting a typical, several years long, AGN light curve, let us say, 
in half changes the measured DRW parameters, because in fact the true underlying stochastic process does not change. 
The importance of the light curve length for DRW were skimmed in \cite{2010ApJ...721.1014M} and
\cite{2016ApJ...826..118K} (and recently recognized in \citealt{2016arXiv161000008F}), but -- as we will show -- were significantly underestimated (or not well understood).
In this paper, we will thoroughly study all plausible secondary effects on the measured DRW parameters and show 
that DRW is not a simple ``black box'' that returns correct parameters.
In particular, we will be interested in how well we are able to estimate the DRW parameters from the ground-based light curves, 
given a range of ratios ($\ri=\tau_{\rm inp}t_{\rm exp}^{-1}$) of the input decorrelation timescale ($\tau_{\rm inp}$) to the experiment length ($t_{\rm exp}$), 
different cadences (2--80 days), light curve lengths (including 4-month-long seasonal gaps), 
and photometric properties of a typical ground-based survey 
(here the Optical Gravitational Lensing Experiment\footnote{{\tt http://ogle.astrouw.edu.pl}, see \cite{2015AcA....65....1U} for details.} (OGLE) and/or the Sloan Digital Sky Survey\footnote{{\tt http://www.sdss.org}, see \cite{2000AJ....120.1579Y} for details.} (SDSS)).
We will perform extensive Monte Carlo simulations and modeling of AGN light curves using DRW to study the biases between the input (hence assumed to be true) and output (measured) model parameters
as a function of the source and survey properties.
In Section~\ref{sec:simulations}, we present the simulation and modeling setup, while the results are discussed in
Section~\ref{sec:discussion}. The paper is summarized in Section~\ref{sec:summary}.


\section{DRW simulations and modeling}
\label{sec:simulations}

We simulate and then model AGN light curves as a DRW stochastic process that is characterized by the covariance matrix of the signal

\begin{equation}
S_{ij} = \sigma^2 \exp{\left(-\frac{|t_i-t_j|}{\tau}\right)},
\label{eqn:sfunc}
\end{equation}
\noindent
where $t_i-t_j=\Delta t$ is the time interval between the $i$th and $j$th epochs, and $\sigma^2$ is the signal variance, while 
its relation to $\hat{\sigma}$ or $SF_\infty$ is $\sigma = \hat{\sigma}\sqrt{\tau/2}$ and $\sigma=SF_\infty/\sqrt{2}$
 (see \citealt{2009ApJ...698..895K,2010ApJ...708..927K,2010ApJ...721.1014M,2011ApJ...728...26M,2012ApJ...753..106M,2011AJ....141...93B,2011ApJ...735...80Z,2013ApJ...765..106Z,2016ApJ...819..122Z}).

\subsection{Light Curve Simulations}

We generate 100-years-long light curves with the cadence of two days. They are later cut into shorter 
time baselines and/or have the cadence reduced, i.e., degraded to the ground-based surveys considered here: 
SDSS Stripe 82, spanning eight years with a median of 60 epochs in the $r$-band,
and the third phase of OGLE (OGLE-III), also spanning eight years with a median of 445 $I$-band epochs in the Large Magellanic Cloud (Figure~\ref{fig:lcs}).
We will also consider 20-years-long OGLE light curves, where we extrapolate the existing cadence into the future.
 
To generate a light curve, the signal chain begins with $s_1=G(\sigma^2)$, where $G(\sigma^2)$ is a Gaussian deviate of
dispersion $\sigma$. The subsequent light curve points are recursively obtained from
\begin{equation}
s_{i+1}=s_i e^{-\Delta t/\tau} + G\left[\sigma^2\left(1-e^{-2\Delta t/\tau}\right)\right],
\label{eqn:nextpoint}
\end{equation}
where $\Delta t = t_{i+1}-t_i$. The observed light curve is obtained from $y_i=s_i+G(n_i^2)+\langle y \rangle$, where $n_i$
is the observational photometric noise, and $\langle y \rangle$ is the mean magnitude (see also \citealt{2010ApJ...708..927K,2011ApJ...735...80Z,2016MNRAS.459.2787K}).

One of the most important quantities is a survey's photometric noise and we adopt it here to be $\sigma_{\rm OGLE}=\sqrt{0.004^2+\exp(1.63\times(I-22.55))}$~mag for OGLE, estimated for
one of the Large Magellanic Cloud fields (\citealt{2008AcA....58...69U,2015AcA....65....1U,2016AcA....66....1S}), while for the SDSS
we adopt $\sigma_{\rm SDSS}=\sqrt{0.013^2+\exp(2.0\times(r-23.36))}$~mag (\citealt{2007AJ....134..973I,2016ApJ...826..118K}), obtained from Stripe 82 AGNs. 
The final simulations' cadence is akin to the original surveys. The exact choice of the input timescales and amplitudes will be described in Discussion.

\subsection{Light Curve Modeling}

The light curves are modeled by means of a method that optimally reconstructs irregularly sampled data, developed by
 \cite{1992ApJ...385..404P,1992ApJ...385..416P} and \cite{1994comp.gas..5004R} (commonly referred to as the ``PRH'' method), and
later adopted in AGN variability studies (\citealt{2010ApJ...708..927K,2011ApJ...735...80Z,2013ApJ...765..106Z,2016ApJ...819..122Z}).

While the full derivation of the method is presented in \cite{2010ApJ...708..927K} and \cite{2011ApJ...735...80Z,2013ApJ...765..106Z,2016ApJ...819..122Z}, we will briefly present its basic concepts here.
An AGN light curve $\bfy(t)$ can be represented as a sum of the true variable AGN signal $\bfs(t)$ (with the covariance matrix $S$), noise $\bfn$ (with the covariance matrix $N$), 
matrix $L$, and a set of linear coefficients $\bfq$ (e.g., that can be used to subtract or add the mean light curve magnitude or any trends), so
$\bfy(t) = \bfs(t) + \bfn + L \bfq$.

The likelihood of the data given $\bfs(t)$, $\bfq$, and model parameters $\tau$ and $\hat{\sigma}$
is (Equation~(A8) in \citealt{2010ApJ...708..927K}, Equation~(11) in \citealt{2013ApJ...765..106Z})
\begin{equation}
   \mathcal{L}\left(\bfy \bigl| \bfs,\bfq,\tau, \hat{\sigma}\right)=|C|^{-1/2}|L^T C^{-1} L|^{-1/2}
   \exp\left( -\frac{ \bfy^T C_\perp^{-1} \bfy}{2} \right),
\end{equation}
where $C=S+N$ is the total covariance matrix of the data,
and $C_\perp^{-1} = C^{-1}-C^{-1}L (L^T C^{-1} L)^{-1} L^T C^{-1}$. To measure the model parameters, the likelihood $\mathcal{L}$ is optimized (maximized) 
in $\sim O(N_{\rm data})$ operations by noting that the inverse of the covariance matrix has a tridiagonal form (\citealt{1994comp.gas..5004R}).

In \cite{2010ApJ...708..927K} and \cite{2010ApJ...721.1014M}, the reported parameters are ``best-fit values'' taken at the likelihood maximum ($\mathcal{L}_{\rm best}$),
and the 1$\sigma$ uncertainties are obtained after projecting the likelihood surface onto 1D for the parameter of interest and taking $\Delta \ln(\mathcal{L}_{\rm best})-0.5$, 
corresponding to $\Delta \chi^2=1$ (right column of Figure~\ref{fig:lcs}). There are two important limits of the DRW model, when $\tau\rightarrow 0$ and $\tau\rightarrow\infty$.
The first one corresponds to a broadening of the photometric error bars, because the covariance matrix of the signal becomes $\sigma^2\delta_{ij}$. We calculate the likelihood
for $\tau\rightarrow 0$, hereafter $\mathcal{L}_{\rm noise}$, and require the best model to return a better fit by $\Delta\chi^2>4$, so $\ln{\mathcal{L}_{\rm best}}>\ln{\mathcal{L}_{\rm noise}}+2.0$.
To avoid unconstrained parameters, \cite{2010ApJ...708..927K} and \cite{2010ApJ...721.1014M} used logarithmic priors on the model parameters
 $P(\tau)=1/\tau$ and $P(\hat{\sigma})=1/\hat{\sigma}$, and we will use the exact same priors throughout this manuscript.

In the {\sc python} implementation of this method by \cite{2011ApJ...735...80Z,2013ApJ...765..106Z,2016ApJ...819..122Z}, the {\sc Javelin}\footnote{\tt https://bitbucket.org/nye17/javelin} software, 
the best solution is optimized with the {\sc aomeba} method and the likelihood maximum is explored with MCMC to find the parameter uncertainties. 
One can choose to use or waive the priors on the model parameters.


\section{Discussion}
\label{sec:discussion}

Of the two DRW model parameters, the amplitude and decorrelation timescale, the latter is of the highest interest as it can
be directly linked to (or interpreted as) one of the characteristic accretion disk timescales: dynamical, thermal, or viscous (e.g., \citealt{2006ASPC..360..265C,2008NewAR..52..253K}).
For a typical black hole mass of $10^8$~$M_\odot$ and the accretion thin disc radius of $0.01$ pc (with $\alpha=0.1$), the dynamical timescale is $\sim$1 year, the thermal timescale is $\sim$10 years,
and the viscous timescale is of order of a million years (e.g., \citealt{2001ApJ...555..775C}). 
It is obvious that the first two timescales are the primary candidates for identification with the DRW variability timescales.

We begin our simulations and discussion with a question: {\it is the DRW model able to return the input parameters in an ideal situation?}
We simulate light curves that are extremely long (20000 days) with 2000 data points, the decorrelation timescale is 200 days (many times shorter than the experiment length) and $SF_\infty=0.20$~mag.
The answer to the asked question is {\it yes}. The likelihood surfaces have narrow peaks and the highest likelihoods are well-matched to the input parameters.

\begin{figure*}
\centering
\includegraphics[width=7.0cm]{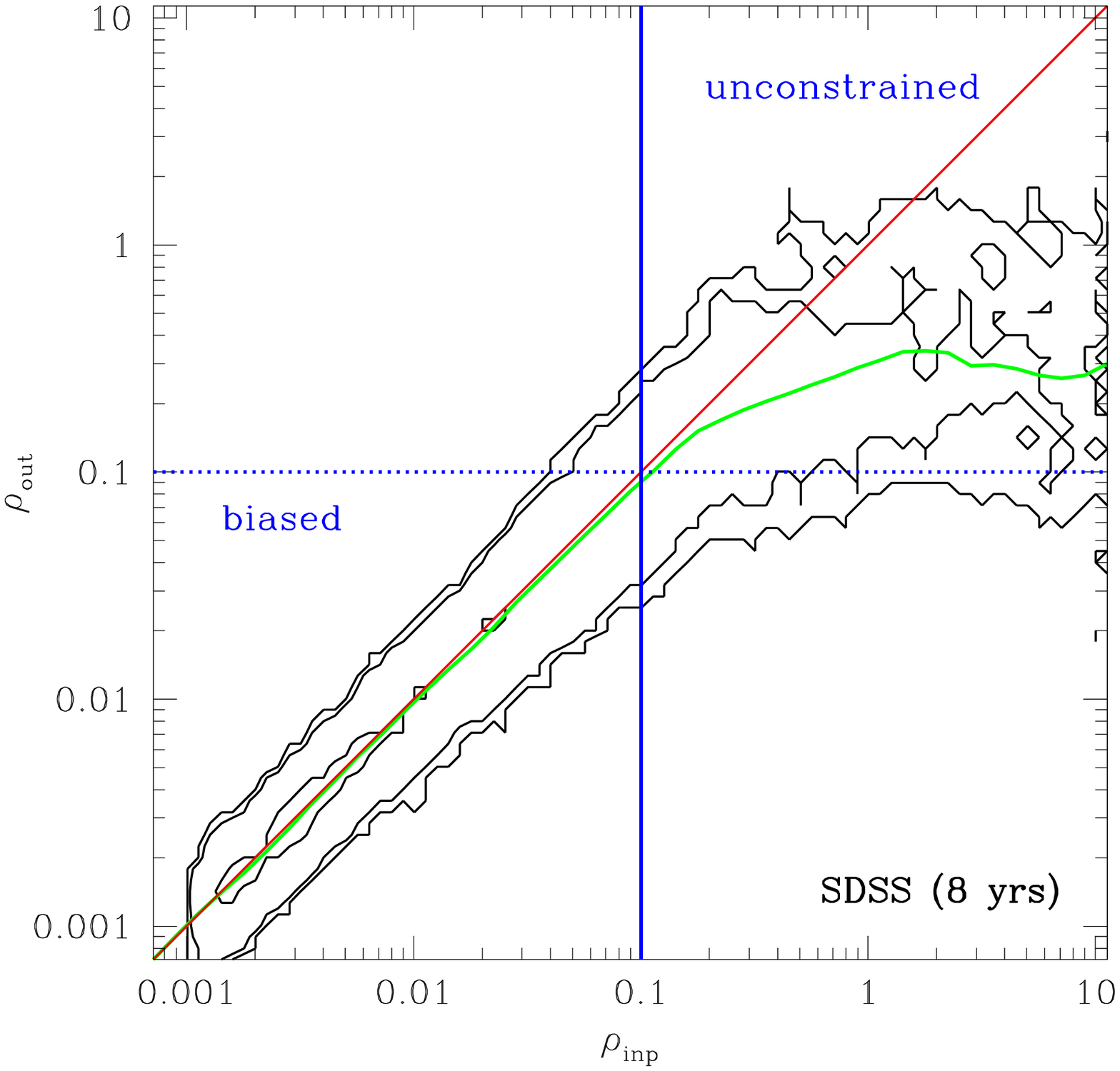} \hspace{0.1cm}
\includegraphics[width=7.0cm]{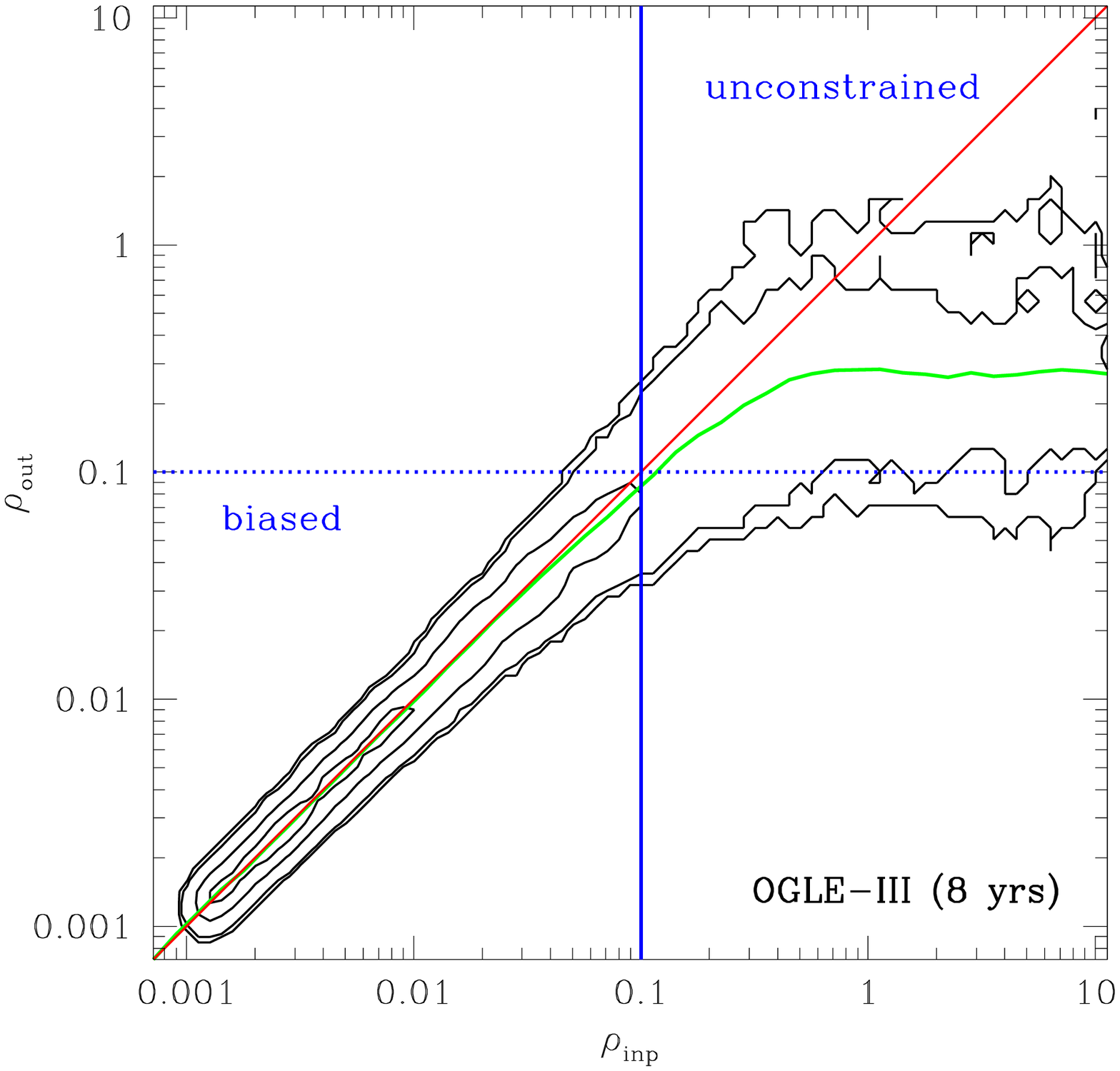}\\
\caption{Bias in the DRW decorrelation timescale caused by the experiment length. 
As contours we present two sets of 10000 AGN light curve realizations modeled as DRW, SDSS (60 epochs; left) and OGLE-III (445 epochs; right),
with $SF_\infty=0.20$~mag and a wide range of input timescales, where $r\approx17$ mag and $I\approx18$~mag, respectively.
AGN light curves that are long compared to the true decorrelation timescale, at least ten times longer ($\ri<0.1$),
are the best source of the correctly measured decorrelation timescales (the ``biased'' region). Light curves that are short and comparable to the decorrelation timescale ($\ri\approx1$) 
will not yield reliable decorrelation timescales and they are unconstrained (the ``unconstrained'' region). 
The typical measured timescale in the ``unconstrained'' region is $\sim$20--30\% of the experiment span, independent of the
input (true) value. The border between the biased and unconstrained regions is $\ri=0.1$. Location of the border line is independent of
the input amplitude $SF_\infty$ and the source magnitude. While not shown, the typical uncertainty on $\ro$ is half of the outer contours.}
\label{fig:bias1}
\end{figure*}

\begin{figure}
\centering
\includegraphics[width=7.5cm]{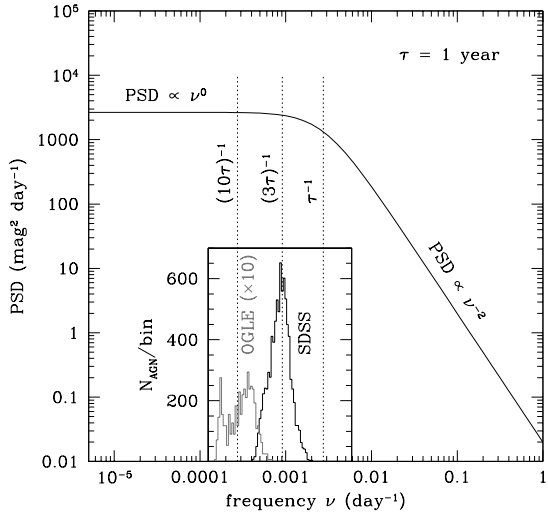}
\caption{Power spectral density (PSD) for DRW. We mark the frequencies $\tau^{-1}$, $(3\tau)^{-1}$, and $(10\tau)^{-1}$. 
The experiment length shorter than $\sim10\tau$ is insufficient to measure $\tau$ because
such light curves simply do not (for $\nu>\tau^{-1}$) or weakly (for $\nu>(10\tau)^{-1}$) probe the white noise part (${\rm PSD} \propto \nu^0$) of the spectrum.
The inset presents the histogram of the SDSS Stripe 82 experiment lengths (hence the longest timescales probed) divided by $(1+z)$ for $\sim$9000 SDSS AGNs and 
500 $I<19$ mag 20-years-long AGN light curves from combined phases of OGLE.
If the true decorrelation timescale is one year then the SDSS S82 data do not probe the white noise part of PSD, 
so cannot constrain the decorrelation timescales.}
\label{fig:power}
\end{figure}

The next important question to answer is: {\it what is the impact of the experiment span on the recovered decorrelation timescales?} 
To answer this question, we simulate 
two sets of 10000 AGN light curves with $SF_\infty=0.20$~mag, one for SDSS ($r=17$~mag; 60 epochs; 8 years) and another for OGLE ($I=18$~mag; 445 epochs; 8 years), 
where the ratio ($\ri=\tau_{\rm inp}t_{\rm exp}^{-1}$) of the input DRW timescale ($\tau_{\rm inp}$) to the experiment length ($t_{\rm exp}$) is $0.001\lesssim\ri\lesssim15$. 
We model these light curves with DRW and find that for output ratios $\ro\geqslant0.1$ (so $\ri\geqslant0.1$), where $\ro=\tau_{\rm out}t_{\rm exp}^{-1}$,
DRW returns incorrect timescales ($\tau_{\rm out}$; the ``unconstrained'' region in Figure~\ref{fig:bias1}), while
for $\ro\leqslant0.1$ (the ``biased'' region) the input and output parameters are closely matched. The name of the ``unconstrained'' region originates from a fact that
regardless of the input timescale, the output timescales have the median of $\sim$20--30\% of the experiment length ($\ro\approx 0.2$--$0.3$; discussed below). 
Reconstruction of the original timescales for AGNs from this region is impossible and
in Figure~\ref{fig:power}, we present the reason: we mark the frequencies $\tau^{-1}$, $(3\tau)^{-1}$, and $(10\tau)^{-1}$ on top of the power spectral density (PSD) for the DRW model.
From this figure one can readily see why the experiment length shorter than $\sim10\tau$ is insufficient to measure $\tau$ -- 
such light curves simply do not or weakly probe the ``flat'' white noise part, which is vital to the DRW timescale determination.
A successful survey for AGN variability should optimally probe the PSD at both low frequencies (long timescales), i.e., the white noise part,
as well as all the high frequencies (short timescales) to constrain the red noise part of PSD.

As a verification, we have checked the dependence of the transition between the biased and unconstrained regions 
on the variability amplitude and source magnitude. We performed additional, nearly identical sets of simulations with 
$SF_\infty=0.10$~mag and $SF_\infty=0.30$~mag, and $I=16$--19 mag. The border is an invariant.

\subsection{The unconstrained region}

We will now consider the two regions separately, starting with the unconstrained region. This region is of particular interest
because typical AGN light curves have lengths of order of the decorrelation timescale, hence they will most likely reside in this region. 
\cite{2016ApJ...826..118K}, using a set of 9000 SDSS $r$-band AGN light curves (\citealt{2007AJ....134..973I,2010ApJ...721.1014M}),
showed that SFs (being model independent quantities) flatten at long timescales, and {\it the rest-frame (ensemble) decorrelation timescale is of about one year}.  
We will now adopt this value for our discussion, regardless of its any plausible correlations with the physical AGN parameters.

The first obvious finding, as already explained in the previous section, is that rest-frame light curves shorter than $\sim10\tau$ ($\ri>0.1$) will provide unreliable, 
so meaningless DRW timescales ($\ro$ in Figure~\ref{fig:bias1}). 
AGNs are typically distant sources and have considerable redshifts $z$. The higher the AGN redshift, the shorter timescales we probe, 
only aggravating the situation, because the experiment length must be even greater, i.e., $10\tau(1+z)$ years long or more.

The DRW timescale from the unconstrained region is a constant fraction of the experiment length 
($\ro$ from the unconstrained region in Fig~\ref{fig:bias1}). The rest-frame experiment length is proportional to $(1+z)^{-1}$, so is the DRW timescale.
From Table~1 in \cite{2010ApJ...721.1014M} we learn that $\tau\propto(1+z)^{-0.7\pm0.5}$, which is an obvious manifestation 
of the experiment length effect, consistent at 0.6$\sigma$ with our interpretation that the data are too short, but also less strongly consistent with no dependence on redshift (at 1.4$\sigma$).

\begin{figure*}
\centering
\includegraphics[width=5.8cm]{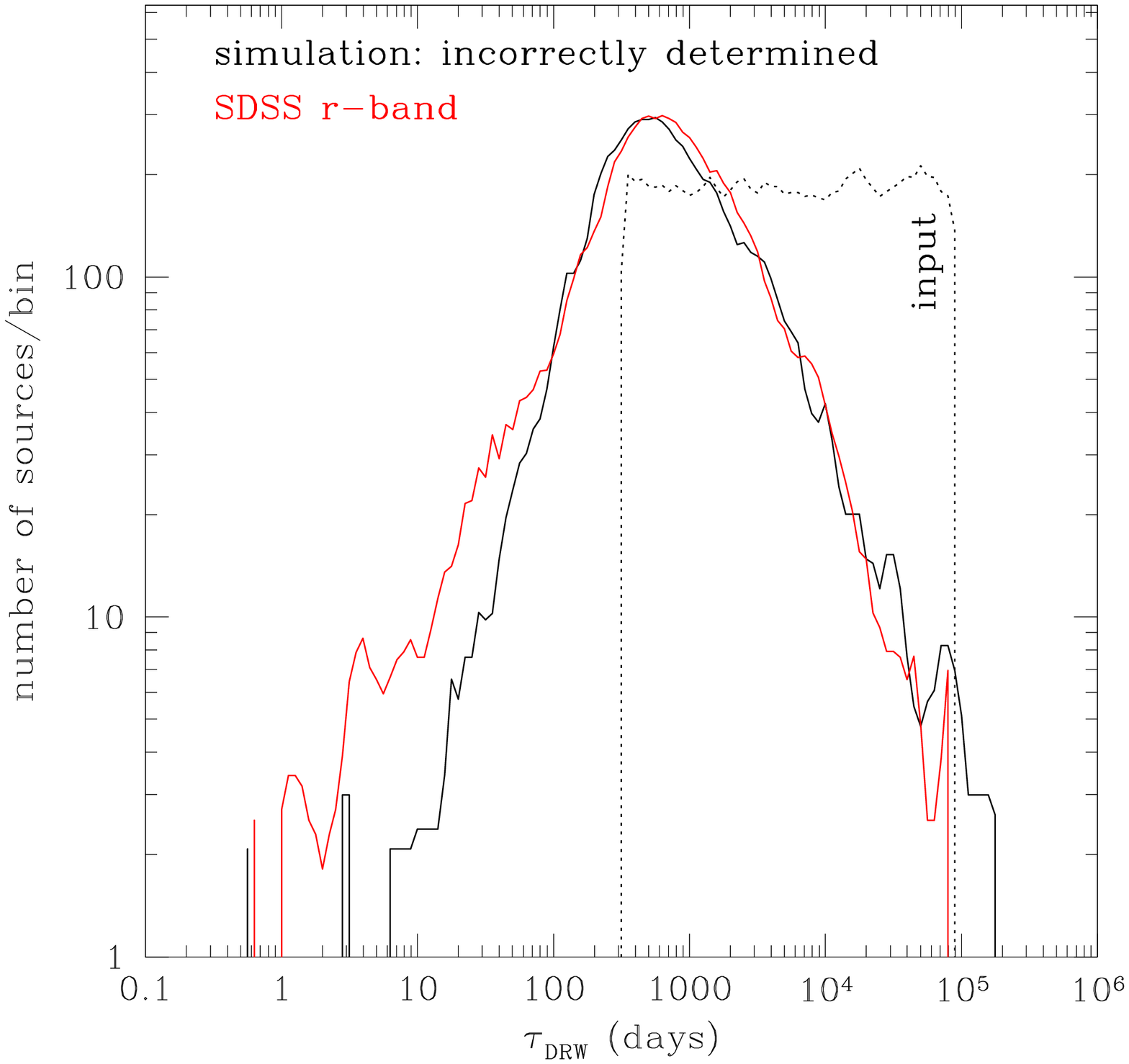} \hspace{0.1cm}
\includegraphics[width=5.8cm]{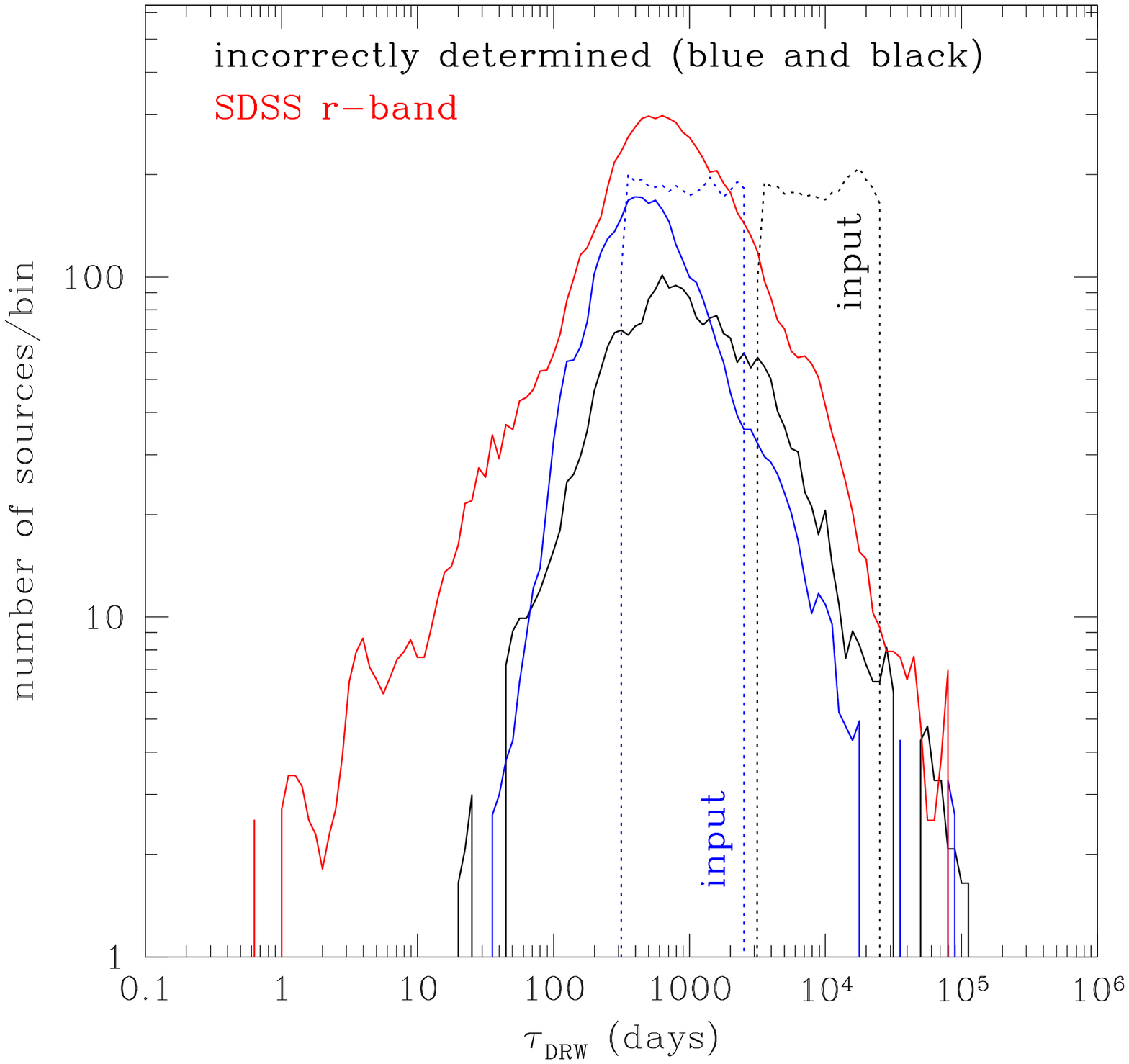}\\
\includegraphics[width=5.8cm]{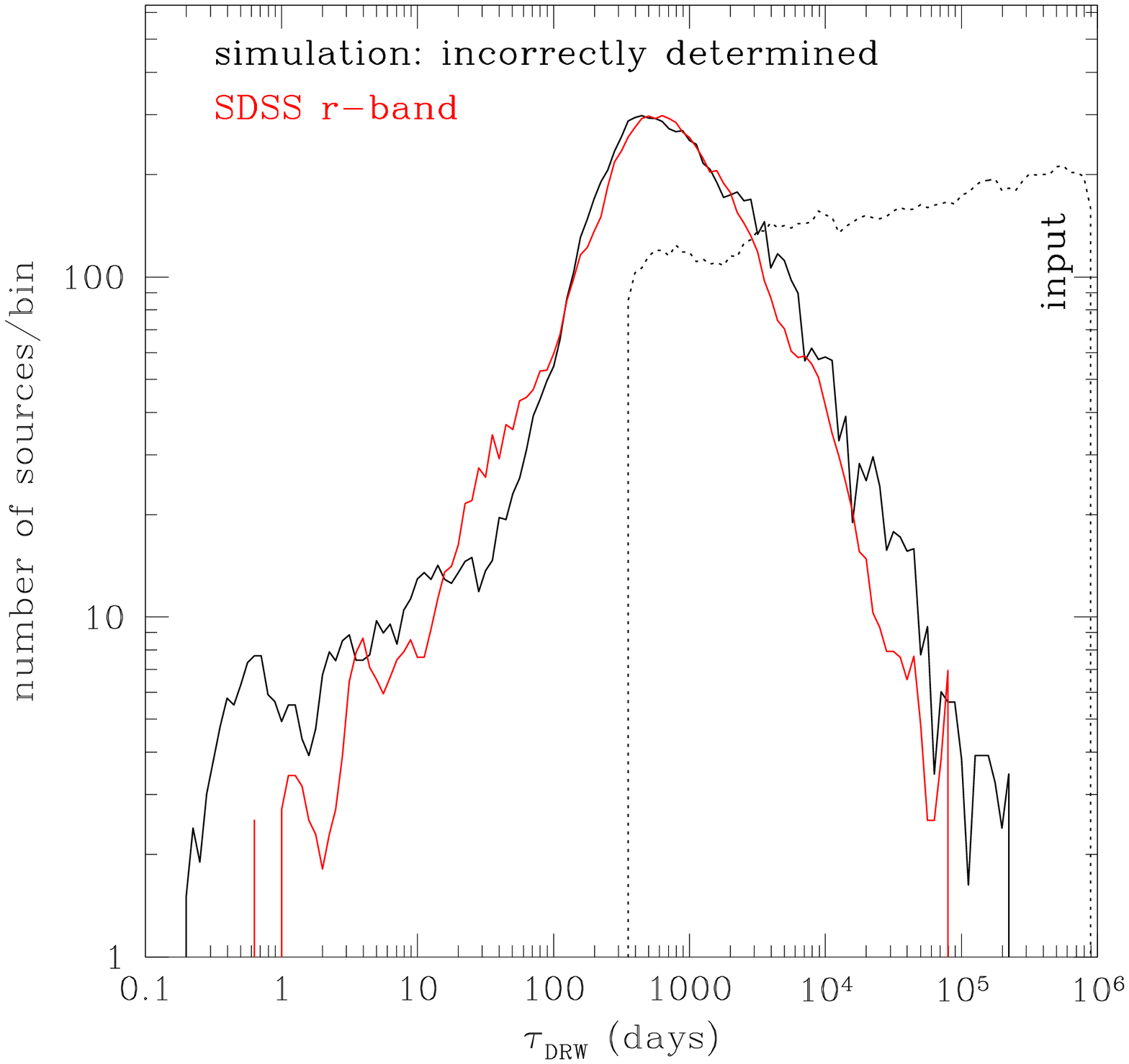} \hspace{0.1cm}
\includegraphics[width=5.8cm]{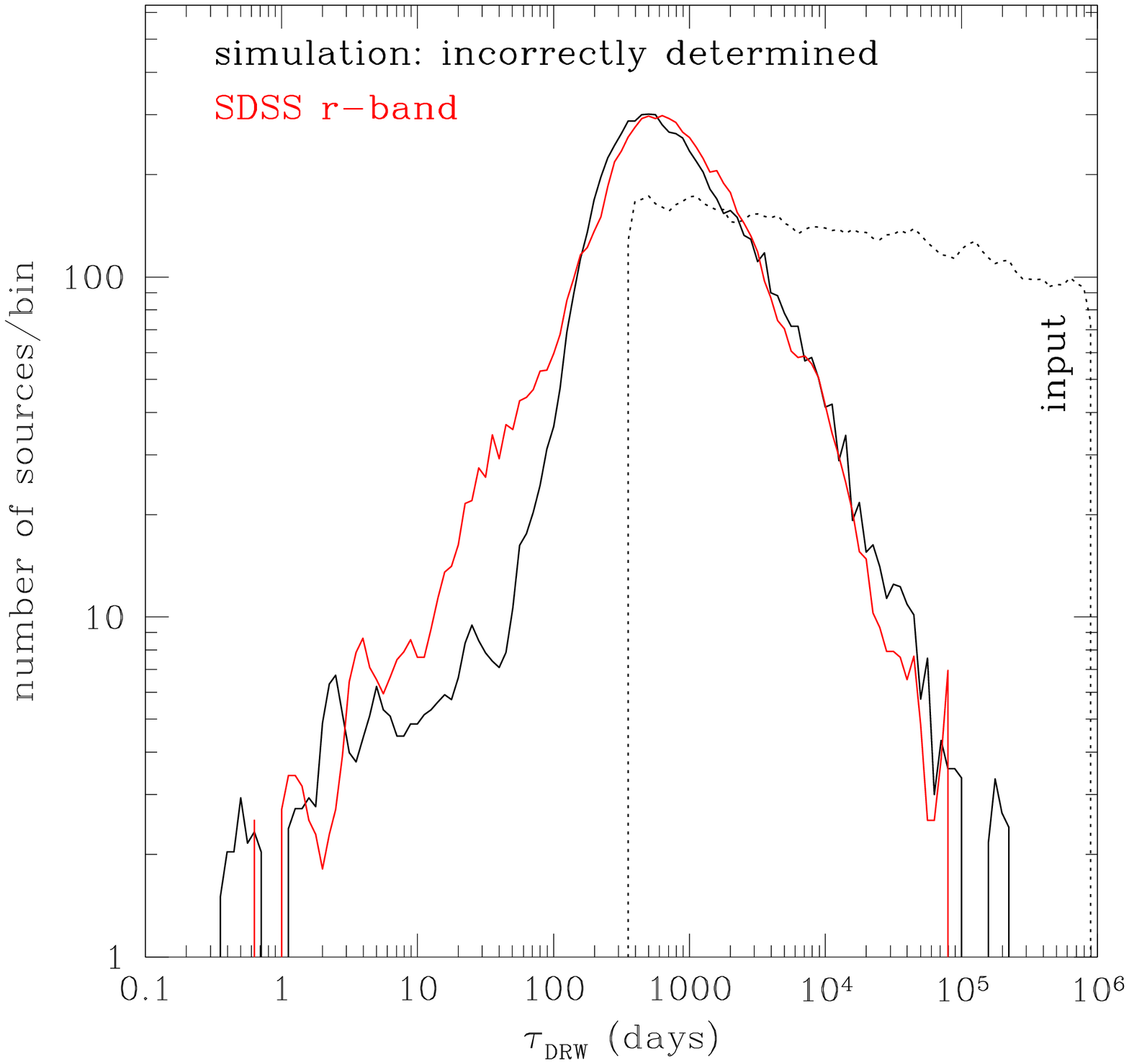}\\
\caption{Simulations and modeling of 10000 SDSS AGN light curves using DRW, where the input timescale is longer than 10\% of the experiment span 
(from the ``unconstrained'' region in Figure~\ref{fig:bias1}).
The input distribution of $\tau$ is marked with dotted lines, while the output (measured) distribution of timescales is marked with the solid black/blue lines. 
The measured distribution of timescales for the real SDSS $r$-band AGN light curves is shown in red. 
It is obvious that DRW is unable to determine true decorrelation timescales if the input timescale is 
longer than 10\% of the experiment span (if it belongs to the ``unconstrained' region in Figure~\ref{fig:bias1}).
A flat input $\tau$ distribution is shown in the top left panel and it is split in two flat distributions in the top right panel. In the bottom row,
we used input distributions that are rising (left) and falling (right) with increasing $\tau$. All the recovered distributions look akin to the SDSS one,
 and for the measured timescales longer than 100 days the KS test identifies the recovered and SDSS $\tau$ distributions as drawn from the same distribution. 
This simply means that the timescales obtained from SDSS (and in the future Gaia/LSST/Pan-STARRS) are unconstrained and should not be used to search for correlations with 
the physical parameters of AGNs, because the true values of $\tau$ and hence their distribution stays unknown.}
\label{fig:bias2}
\end{figure*}

{\it How does the distribution of the recovered decorrelation timescales from the ``unconstrained'' region look like?}
Figure~\ref{fig:bias2} answers this question graphically. We will consider several input timescale distributions for the SDSS survey: flat, increasing and decreasing number of AGNs per timescale bin.
The recovered distributions of timescales are distant from the input ones, but they closely resemble one another. 
In other words, from the recovered distribution we are unable to guess what the input distribution was. It could have been either flat, or could have had other shapes.
It is also unclear what the highest ratio could have been, because continuous distributions of ratios as high as 10 or 100, deliver nearly identical final distributions.

The most notable finding is made when comparing these recovered SDSS distributions to the real SDSS timescale distribution. They are nearly identical for the timescales longer than 100 days and
a two-sample Kolmogorov-Smirnov (KS) test returns the significance level $P=0.67$ for the null hypothesis that the data sets are drawn from the same distribution 
for 81 points in both distributions (for the top left panel in Figure~\ref{fig:bias2}). 
The bottom line here is that the true SDSS timescale distribution is unknown and it could have been for example flat. 
The unconstrained SDSS timescales have been used to search for their correlations with the physical AGN parameters such as the black hole mass, luminosity, Eddington ratio, or wavelength. 
As shown in Figure~\ref{fig:bias2}, the posterior timescale distribution has nothing to do with the input one, hence it is truly meaningless, and so are the reported correlations.

We provide ``a rule of thumb'' for identification of projects having AGN light curves, and hence the DRW parameters inside the ``unconstrained'' region. The median recovered timescale from this 
region is $\sim$20-30\% of the experiment span, regardless of the true timescale. So for example the SDSS Stripe 82 experiment lasted 2920 days (eight years),
it will be unable to correctly identify timescales longer than $\sim$300 days or 0.8 year at $z=0$ (greater than 10\% of the experiment length), 
and then they will typically have the measured value of $600\lesssim\tau_{\rm out}\lesssim900$ days (20--30\% of the experiment length). In fact the SDSS distribution peaks at 560 days, the median is 617 days. This clearly confirms inability of obtaining correct DRW timescales from SDSS. Unfortunately, it appears that the same will be the case for most past, currently existing, 
or near-future surveys that rarely are planned to last a decade or longer.
There is clearly a possibility for correctly measuring the decorrelation timescale from the OGLE survey. There is about 800 AGNs observed since 2001 (16-years-long as of now)
and a smaller subsample of AGNs observed since 1997 (20-years-long; see inset of Figure~\ref{fig:power}). If the true decorrelation timescale is indeed of about a year, this means
that for AGNs with $z<0.6$ the rest-frame light curves lengths will be $\sim$10 years. There exist 115 AGNs with at least 16-years-long light curves, $z<0.6$, and $I<19.5$~mag 
(the photometric noise 40\% of the expected AGN variability), discovered mostly (102)
by the Magellanic Quasars Survey (\citealt{2011ApJS..194...22K,2012ApJ...746...27K,2013ApJ...775...92K}) and a smaller number (13)
from other works (\citealt{2002ApJ...569L..15D,2003AJ....125.1330D,2003AJ....126..734D,2005A&A...442..495D,2003AJ....125....1G}).

\begin{figure*}
\centering
\includegraphics[width=14.0cm]{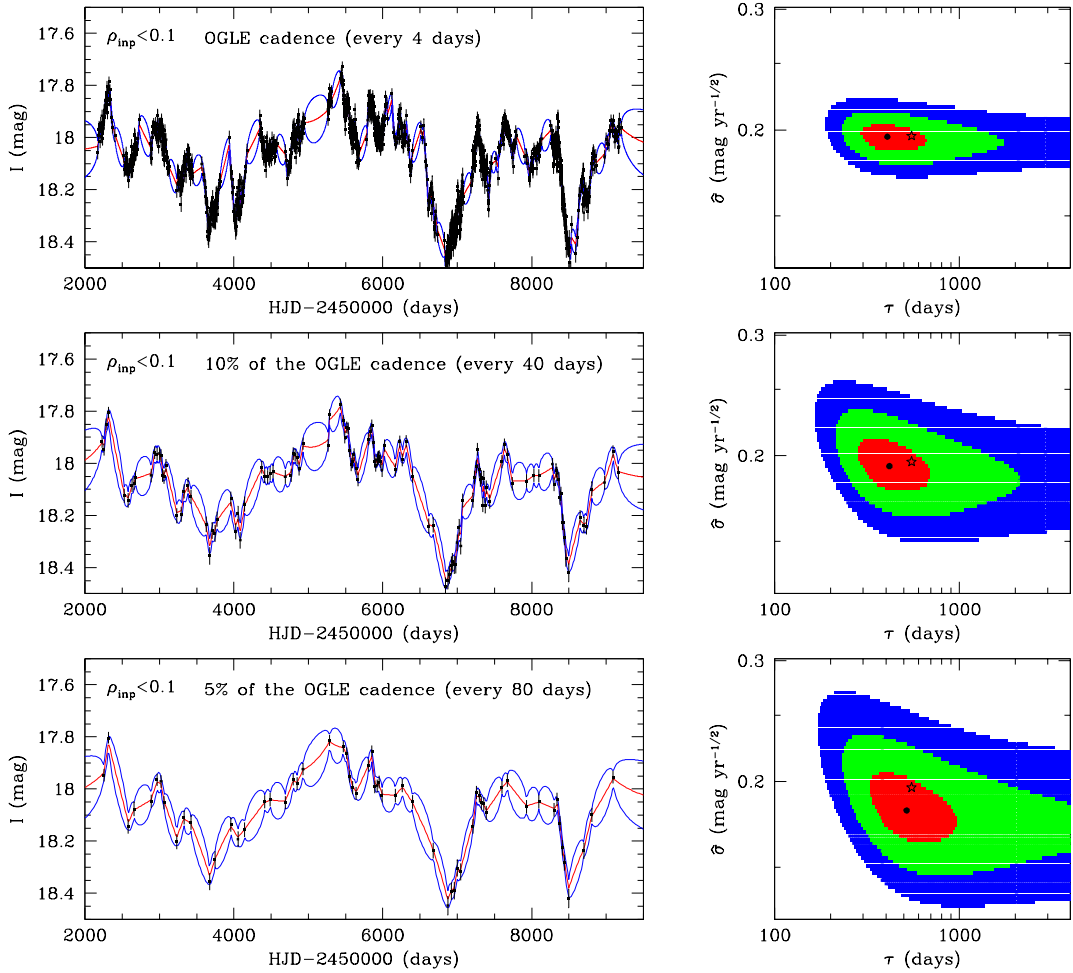}
\caption{Left column: Simulated OGLE AGN light curves spanning 20 years, where we extrapolated the cadence into the future (top). We assumed the AGN to be at $z=0.5$ and $\tau=1$ year,
so $\ri=0.07$ (smaller than the limiting $\ri=0.1$). 
The typical OGLE cadence during observing seasons of LMC is about four days. In the middle and bottom panels, we keep only every 10th and 20th point from the upper light curve. 
The red line is the best-fit DRW model and the blue lines present $1\sigma$ ``error snakes''. The right column presents the model likelihood surface 
for the three light curves. The black point is the maximum likelihood point, while the red, green, and blue areas represent $\Delta\ln{\mathcal{L}}=-0.5$, $-2.0$, and $-4.5$,
corresponding to $\Delta\chi^2=1$, 4, and 9 (or 1, 2, and 3$\sigma$), respectively. The open star marks the input parameters.
In all three cases the timescale is measured with similar precision and bias, while the
uncertainty in the variability amplitude increases, because the middle and bottom light curves do not probe the short timescales.}
\label{fig:cad}
\end{figure*}

\cite{2016ApJ...826..118K} using optical SFs shows that $\tau \propto M^{0.4}$ (much weaker dependence than $\tau\propto M^{2.2}$ from \citealt{2006Natur.444..730M} from X-rays), 
meaning that less massive black holes will have the timescales of several months only. It is then plausible to derive meaningful DRW parameters for low redshift,
low mass black holes from about a decade long existing surveys, provided the cadence is sufficient, i.e., the observing cadence is better than the DRW timescale.
These objects will appear in the ``biased'' region of Figure~\ref{fig:bias1}. 

\subsection{The ``biased'' region and dependencies on other variables}

We will be now interested in reconstructing the DRW parameters as a function of several survey's variables,
that include the average light curve magnitude, length, sampling (cadence), and additional host light for AGNs from the ``biased'' region of Figure~\ref{fig:bias1}.

Similarly to the analysis of the unconstrained region, we perform Monte Carlo simulations of AGN light curves 
for the asymptotic amplitude $SF_\infty=0.20$~mag and the decorrelation timescale of $\tau=1$ year.

From Figure~\ref{fig:bias1}, we see that the typical measured timescale asymptotically approaches the true value with the increasing experiment length (with decreasing $\rho$).

In a large simulation set probing cadences 2--80 days and a wide magnitude range, we study the impact of data sampling on the derived DRW parameters. 
We find that there is nearly no influence of cadence on the timescale, provided that the sampling is still better than the decorrelation timescale.
In Figure~\ref{fig:cad}, we present a visualization for three example 
cases, where we extrapolate OGLE sampling into the future to obtain 20-years-long light curves.
We then degrade the cadence to 10\% and 5\% of the original and we assumed an AGN to be at $z=0.5$, hence $\tau_{\rm inp}=1(1+z)$ year,
what gives $\ri\approx0.07$. It is obvious that the parameter uncertainties (right column of Figure~\ref{fig:cad})
are similar for the timescale (nearly independent of cadence), while the amplitude uncertainty increases for sparser samplings.
This is because there is less information on the high frequencies in PSD than in the original light curve.
The bottom line is that {\it we observe minute dependence of the survey's cadence on the measured timescale for $\ri<0.1$}. 

An AGN light curve may contain not only the AGN light, but also additional light coming either from the AGN's host galaxy or other objects near and/or along the line-of-sight.
We define blending of 20\% if 80\% of the light comes from an AGN and 20\% from another source. 
We perform 1000 Monte Carlo simulations for blending spanning 0--100\% for several magnitude levels.
Blending seems to weakly affect the recovered timescales,
while the modified amplitude is (unsurprisingly) strongly affected by blending. The ratio of the recovered-to-input
variability amplitude is directly proportional to the ratio of the AGN-to-the-total light.
The goodness of fit seems to weakly improve for the increasing blending, most likely because it is easier to fit a diluted variability (or lack of it) than
a full amplitude variability. The problem of AGN host light may be alleviated by using differential flux light curves.


\section{Summary}
\label{sec:summary}

A simple procedure of cutting a typical AGN light curve in half and then modeling both the full and short ones with DRW, has returned different variability parameters.
This worrisome discovery has triggered us to study the impact of the experiment length, cadence, and photometric quality on the returned DRW model parameters. 

The most important finding here is that to correctly measure the DRW timescale (the only parameter that can be directly linked to AGN physics), 
the experiment length must be at least ten times longer than the true signal decorrelation timescale (rest-frame). 
Otherwise, DRW is unable to correctly recognize the timescale, simply because the light curves do not probe the pink and white noise parts of PSD. 
We provide a rule of thumb for identification of too short AGN light curves to be modeled by DRW. If the typical measured decorrelation timescale is $\sim$20--30\% of the experiment length
one can be sure the timescale is measured incorrectly. 
This is also the case for SDSS, so the validity of correlations between DRW variability parameters and the physical AGN parameters
from \cite{2010ApJ...721.1014M} is refuted. In particular, these authors find that $\tau\propto(1+z)^{-0.7\pm0.5}$, which is an obvious manifestation 
of the experiment length effect on the DRW parameters, 
and we show the expected correlation for too short data sets to use DRW is $\tau\propto(1+z)^{-1}$.
At present usability of DRW to provide AGN variability parameters is questioned (please also note, we used priors on the DRW parameters 
($1/\tau$ and $1/\hat{\sigma}$) that prevent $\tau\rightarrow\infty$) with two exceptions: 
(1) Because DRW is based on a method that optimally reconstructs 
``missing'' data, it still has a lot of potential in interpolating AGN light curves. One has to guess, however, what the correct DRW parameters are
for such an interpolation, and they can be obtained from SFs (\citealt{2016ApJ...826..118K}).
(2) Another avenue for measuring the DRW parameters is to analyze both low redshift and the least massive AGNs, as \cite{2016ApJ...826..118K} using optical SFs
shows that $\tau \propto M^{0.4}$, meaning that at the light weight end of the BH mass spectrum, the timescales should be of several months, so 
measurable from currently existing surveys.

Most past, existing, and near-future surveys, unfortunately, are/will be unable to measure DRW timescales for majority of AGNs, unless
they span a vicinity of two decades or longer, even with a sparse sampling.
The first credible decorrelation timescales for individual AGNs may come from the OGLE survey,
being 16-years-long for 800 AGNs (or 20-years-long for a small subsample) and continuously growing.  

It seems that given their planned lengths none of the surveys such as Gaia, LSST, or Pan-STARRS will have the 
sufficient length to measure the decorrelation timescales for typical AGNs correctly. They will, however, be able to probe the covariance matrix of the signal $S_{ij}$
which can be obtained from SF as 
$SF(t_i-t_j)= \sqrt{2\sigma_n^2+2\sigma_s^2-2S_{ij}}$, where $\sigma_s^2$ and $\sigma_n^2$ are the signal and noise variances, respectively (see \citealt{2016ApJ...826..118K} for a review).
Given a large number of AGNs monitored, these surveys will be able to trace any plausible correlations of the covariance matrix shape (via the SF slope) 
with the luminosity, black hole mass, and Eddington ratio.

An intriguing finding we made was that for sufficiently long  AGN light curves ($>10$ years rest-frame),
the cadence of a survey seems nearly unimportant for determination of the DRW decorrelation timescale of $\tau=1$ year 
(at least in a considered range of 2--80 days) but the light curve sampling must be still shorter than the timescale (but high cadence is not necessary).
The important lesson is that a light curve should optimally probe PSD at both red, pink and white noise parts.

Another DRW bias was found recently by \cite{2016MNRAS.459.2787K}. He found a degeneracy in DRW modeling of light curves,
where both non-DRW and DRW light curves are equally well-modeled by the DRW method described in Section~\ref{sec:simulations}, 
however, the former ones at a price of biased parameters.
Because DRW modeling uses an exponential covariance matrix of the signal that translates into the power-law SF slope
of 0.5 or the PSD slope of $-2$ (for $\Delta t<1$~yr; \citealt{2016ApJ...826..118K}), one should make sure
that the underlying variability in a light curve is well-described by a DRW process, prior to the DRW modeling.
It appears that optical light curves of AGNs seem to have PSD slopes $\alpha\lesssim-2$ or SF slopes $\gamma\gtrsim0.5$ 
(\citealt{2011ApJ...743L..12M,2015MNRAS.451.4328K,2016A&A...585A.129S,2016ApJ...826..118K,2016arXiv161103082C}), 
this means that modeling AGN light curves with DRW (having fixed $\alpha\equiv-2$ or $\gamma\equiv0.5$) will produce biased (longer) decorrelation timescales.

\section*{Acknowledgements}

This work has been supported by the Polish National Science Center grants:
OPUS 2014/15/B/ST9/00093 and MAESTRO 2014/14/A/ST9/0012.




\label{lastpage}
\end{document}